\documentclass[aps,pre,preprint,groupedaddress,amssymb]{revtex4-2}
\usepackage{graphicx} 
\usepackage{natbib}
\usepackage{dcolumn}
\usepackage{bm}
\usepackage{amsmath}
\usepackage{epsfig}
\usepackage{color,hyperref}

\begin{document}
\title{Statistical properties of stochastic functionals under general resetting}

\author{V. M\'endez}
\email{vicenc.mendez@uab.cat}
\author{R. Flaquer-Galm\'es}

\affiliation{Grup de F\'{\i}sica Estad\'{\i}stica, Departament de F\'{\i}sica. Facultat de Ci\`{e}ncies, Universitat Aut\`{o}noma de Barcelona, 08193 Barcelona, Spain}

\begin{abstract}
We derive the characteristic function of stochastic functionals of a random walk whose position is reset to the origin at random times drawn from a general probability distribution. We analyze the long-time behavior and obtain the temporal scaling of the first two moments of any stochastic functional of the random walk when the resetting time distribution exhibits a power-law tail. When the resetting times PDF has finite moments, the probability density of any functional converges to a delta function centered at its mean, indicating an ergodic phase. We explicitly examine the case of the half-occupation time and derive the ergodicity breaking parameter, the first two moments, and the limiting distribution when the resetting time distribution follows a power-law tail, for both Brownian and subdiffusive random walks. We characterize the three different shapes of the limiting distribution as a function of the exponent of the resetting distribution. Our theoretical findings are supported by Monte Carlo simulations, which show excellent agreement with the analytical results.
\end{abstract}

\maketitle

\section{Introduction}
In recent years, stochastic resetting has emerged as a powerful concept in the study of random processes, attracting growing attention across various disciplines. This mechanism, which intermittently interrupts the evolution of a stochastic process by returning it to a predefined state, has been shown to significantly alter fundamental properties of the underlying random walk. Originally proposed in the context of Brownian motion \cite{EvMa11,Ev20} stochastic resetting has since been extended to a wide range of processes, including L\'evy flights \cite{Zb24,RaCr24,Ma22}, run-and-tumble dynamics \cite{Br20}, and subdiffusive motion \cite{MeMa21}. Notably, resetting can lead to the formation of a non-equilibrium stationary state and optimize search strategies by minimizing the mean first passage time under certain conditions. This effect is due to the fact that stochastic resetting forces the walker to revisit a point or a region \cite{MeFlCa24}. A similar effect is observed when the walker moves in a heterogeneous media in which the diffusion coefficient depends on space as power law \cite{LeBa19} or when the walker is under the effect of a force field \cite{MeBo22,Evans_2013}, the system reaches a steady state and the mean first passage is finite. Most of the studies deal with a Poissonian resetting process, i.e., when the reset times are distributed according to an exponential distribution. Only a few studies deal with reset times distributed according to general distributions. If this distribution decays as a power-law, then the existence of a steady state and a finite first passage time depend on the characteristic exponent of the distribution and the properties of the underlying random walk \cite{EuMe16,ChSo18,MaCaMe19,ShSo22,BoSo20}. 

A different problem involving resetting consists of studying the statistical properties of stochastic functionals of a random walker when its position is forced to start anew at random times. A prototypical example is the Brownian functional, defined as the time-integral over the trajectory of a Brownian walker \cite{Ma05}. More recently, the case of subdiffusive walkers has been considered as well \cite{TuCaBa09,Ca10}. Among the most widely studied functionals are the occupation time in an interval and the half-occupation time, which have diverse applications in physics, mathematics, and related fields \cite{GoLu01}. Most existing studies regarding stochastic functionals under resetting consider it as Poissonian process \cite{MeSaTo15,Ho19}, and less attention is paid when the resetting times are distributed according to general probability density functions. However, we have recently studied the statistical properties of the occupation time in an interval for a Brownian particle under reset times sampled from a power-law probability density function (PDF) \cite{BaFlMe23}. Despite this progress, a comprehensive theory describing the behavior of general stochastic functionals under arbitrary resetting schemes is still lacking.

In this work, we explore how the statistical properties of stochastic functionals are affected when the position of the underlying random walk is reset to the origin after random times sampled from a general PDF. Since the resetting mechanism is a renewal process, we are able to find an equation that relates the characteristic function of the functional under resetting in terms of two expressions: the resetting times PDF and the characteristic function of the functional without resetting. We find that when the moments of the resetting PDF are finite, the PDF of the functional converges to a Dirac function centered at the mean value in the long-time limit. This is the ergodic case. 

On the other hand, when the resetting times PDF has diverging moments, the characteristic function must be computed on a case-by-case basis, specifying each functional individually.
We further explore the temporal behavior of the first two moments of the functional, which is also useful to study the ergodic properties. We particularize our general results to the half occupation time and obtain its limiting PDF when the resetting times PDF has a power-law tail. More specifically, we obtain the limiting PDF for both normal and subdiffusive underlying random walks.

The paper is structured as follows: In Sec. II we introduce the definition of stochastic functional and its characteristic function (or moment-generating function). In Sec. III we derive the characteristic function of a functional under general resetting, obtain the expression for the first two moments, and illustrate how to analyze the ergodic properties. 
In Sec. IV we prove general results for any stochastic functional. In particular, we show that any stochastic functional becomes ergodic i.e., deterministic in the long-time limit, when the resetting PDF has a finite moments. Likewise, we derive the long-time scaling temporal dependence for the first two moments of any functional when the resetting PDF decays as a power law. In Sec. V we apply our results to the particular case of the half occupation time. We conclude in Sec. VI

\section{Stochastic functionals}
A stochastic functional is defined as the integral up to the measurement time $t$ of a positive function $U[x(\tau)]$ of the trajectory $\{x(\tau); 0\leq \tau\leq t,x(0)=x_0\}$ of a random walker. The stochastic functional is then given by
\begin{equation}
    Z(t|x_0)=\int_{0}^{t}U[x(\tau)]d\tau.
    \label{Z}
\end{equation}
Let $P(Z,t|x_0)$ be the one-time PDF of $Z(t|x_0)$ and $Q(p,t|x_0)$ its characteristic function (or moment-generating function), which is nothing but its Laplace transform with respect to $Z$, this is, 
\begin{eqnarray}
    Q(p,t|x_0)&=&\mathcal{L}[P(Z,t|x_0)]=\int_{0}^{\infty}e^{-pZ}P(Z,t|x_0)dZ\nonumber\\
    &=&\left\langle e^{-pZ(t|x_0)}\right\rangle 
    =\left\langle e^{-p\int_{0}^{t}U[x(\tau)]d\tau}\right\rangle .
    \label{mgf0}
\end{eqnarray}
which satisfies a backward master equation known as the Feynman-Kac equation \cite{Kac49,Kac51}. 
Expanding the exponential in \eqref{mgf0} as a power series, we find
\begin{equation}
    Q(p,t|x_0)=\left\langle \sum_{n=0}^{\infty}\frac{(-p)^{n}}{n!}Z(t|x_0)^{n}\right\rangle =\sum_{n=0}^{\infty}\frac{(-p)^{n}}{n!}\left\langle Z(t|x_0)^{n}\right\rangle 
    \label{mgf}
\end{equation}
Introducing another Laplace transform conjugate to time as
\begin{eqnarray}
  \tilde{Q}(p,s|x_0)=\mathcal{L}[Q(p,t|x_0)]
  =\int_0 ^\infty dt~ e^{-st}Q(p,t|x_0)  ,
  \label{Qps}
\end{eqnarray}
the moments of $Z (t|x_0)$ can be obtained from the derivatives of $\tilde{Q}(p,s|x_0)$  in a systematic manner. To do this, we first define the Laplace transform of the moments 
\begin{align}
    \left\langle \tilde{Z}(s|x_0)^{n}\right\rangle &\equiv\mathcal{L}[\left\langle Z(t|x_0)^{n}\right\rangle ] \nonumber \\
   & = \int_0^\infty dt ~e^{-st} \int_{0}^{\infty}Z(t)^{n}P(Z,t|x_0)dZ.
\end{align}
Making use of Eq. \eqref{mgf0}, it is straightforward to show that the moments are represented in terms of the generating function in the following way
\begin{align}
    \left\langle \tilde{Z}(s|x_0)^{n}\right\rangle =(-1)^{n} \frac{\partial^{n}\tilde{Q}(p,s|x_0)}{\partial p^{n}}\bigg|_{p=0}. 
    \label{moments}
\end{align}
This equation will be used later to compute the moments of the functional once the characteristic function is known.

\section{Characteristic function under resetting}
The goal is to compute the characteristic function of a stochastic functional $Z(t|x_0)_r$ when the underlying random walk is subjected to resetting its initial condition, $x=x_0$.
Consider now that there are $N$ resets over the time interval $[0,t]$. If we define the sequence of time intervals between resets $\{\tau_1, \tau_2, ..., \tau_N, \tau_{N+1}\}$ where $\tau_{N+1}$ is the time since the last reset then
$$
t=\sum_{l=1}^{N+1}\tau_i.
$$
In consequence, the integral in Eq. \eqref{Z} can be expressed as
\begin{eqnarray}
Z(t|x_0)_r=\int_{0}^{\tau_{1}}U[x(t')]dt'+\int_{\tau_{1}}^{\tau_{1}+\tau_{2}}U[x(t')]dt'+...=\sum_{l=1}^{N+1}\int_{\delta_{l}}^{\tau_{l}+\delta_{l}}U[x(t')]dt'
\end{eqnarray}
where $\tau_0=0$ and $\delta_{l-1}=\sum_{j=1}^{l}\tau_{j-1}$. 
Let us define $\varphi (\tau)$ as the PDF of the time elapsed between two consecutive resets. Then, $\tau_i$ with $i=1,...,N+1$ are random variables with PDF $\varphi (\tau_i)$. The probability of no reset until time $\tau$ is
$$
\varphi ^* (\tau)=1-\int_0^\tau \varphi (\tau')d\tau'=\int_\tau^\infty \varphi (\tau')d\tau'.
$$ 
Because of the additivity of $Z(t|x_0)_r$, it is clear that its characteristic function $Q(p,t|x_{0})_{r}$ can be decomposed, when conditioned on these $N$ resettings, into a product of generating functions $Q(p,\tau_i|x_{0})_0$ involving a stochastic trajectory of a random walk between resettings. Hence, $Q(p,t|x_{0})_{0}$ is the characteristic function of $Z$ when the underlying random walk is free of resetting.
Then, the characteristic function $Q(p,t|x_0)_r$ of the functional $Z(t|x_0)_r$,  can be written in the form of a renewal equation \cite{MeSaTo15}
\begin{eqnarray}
Q(p,t|x_{0})_{r}=\sum_{N=0}^{\infty}\Phi_{N}\left(p,t\right)
\label{Qr}
\end{eqnarray}
where
$$
\Phi_{N}\left(p,t\right)=\left[\prod_{l=1}^{N}\int_{0}^{t}F(p,\tau_{l})d\tau_{l}\right]\int_{0}^{t}F^{*}(p,\tau_{N+1})\delta\left(t-\sum_{l=1}^{N+1}\tau_{i}\right)d\tau_{N+1}
$$
and we have defined $F(p,\tau_l)=\varphi(\tau_l)Q(p,\tau_l|x_{0})_0$ and $F^{*}(p,\tau_l)=\varphi^{*}(\tau_l)Q(p,\tau_l|x_{0})_0$.
A standard method for solving such an equation is to use the Laplace transform with respect to the variable $t$. For $N=1$, 
$$
\Phi_{1}\left(p,t\right)=\int_{0}^{t}d\tau_{1}F(p,\tau_{1})\int_{0}^{t}d\tau_{2}F^{*}(p,\tau_{2})\delta(t-\tau_{1}-\tau_{2})
$$ 
whose Laplace transform is
\begin{eqnarray}
 \Phi_{1}\left(p,s\right)&=&\mathcal{L}\left[\Phi_{1}\left(p,t\right)\right]=\int_{0}^{\infty}e^{-st}\Phi_{1}\left(p,t\right)dt\\
 &=&\int_{0}^{\infty}dt\;e^{-st}\left[\int_{0}^{t}d\tau_{1}F(p,\tau_{1})F^{*}(p,t-\tau_{1})\right]=F(p,s)F^{*}(p,s),   
\end{eqnarray}
once the convolution theorem is used. Analogously, for $N=2$ one has $\Phi_2(p,s)=F(p,s)^2F^{*}(p,s)$ and so on. Then, taking the Laplace transform of \eqref{Qr} with respect to $t$, we have
$$
\tilde{Q}(p,s|x_{0})_{r}=\sum_{N=0}^{\infty}\Phi_{N}\left(p,s\right)=\sum_{N=0}^{\infty}F(p,s)^{N}F^{*}(p,s)=\frac{F^{*}(p,s)}{1-F^{}(p,s)}
$$
so that
\begin{eqnarray}
    \tilde{Q}(p,s|x_{0})_{r}=\frac{\mathcal{L}\left[\varphi^{*}(t)Q(p,t|x_{0})_0\right]}{1-\mathcal{L}\left[\varphi(t)Q(p,t|x_{0})_0\right]}.
    \label{Qrlt}
\end{eqnarray}
This equation allows to find the characteristic function of a functional of a random walk under resetting in terms of the characteristic function of the functional of a random walk without resetting. Note that in the absence of resetting $\varphi \to 0$ and $\varphi^*\to 1$, so that $\tilde{Q}(p,s|x_{0})_{r}=\tilde{Q}(p,s|x_{0})_{0}$.

When the resetting mechanism is a Poisson process, the times between resetting events are exponentially distributed with the constant rate $r$. Then, considering the particular case $\varphi(t)=re^{-rt}$, Eq. \eqref{Qrlt} turns out to be
\begin{eqnarray}
    \tilde{Q}(p,s|x_{0})_{r}=\frac{\tilde{Q}(p,s+r|x_{0})_0}{1-r\tilde{Q}(p,s+r|x_{0})_0}
\end{eqnarray}
which is a result derived previously in Refs. \cite{MeSaTo15,Ho19}.

\subsection{Moments}
Next, we want to relate the moments of the functional in the presence of resetting with the moments of the functional without resetting. To do this, we start with Eq. \eqref{Qrlt}. Taking the first derivative of Eq. \eqref{Qrlt} with respect to $p$ we find from \eqref{moments}
\begin{eqnarray}
    \left\langle \tilde{Z}(s|x_{0})\right\rangle _r&=&-\frac{\partial \tilde{Q}(p,s|x_{0})_r}{\partial p^{}}\bigg|_{p=0}=-\frac{\partial}{\partial p^{}}\left(\frac{F^{*}(p,s)}{1-F^{}(p,s)}\right)_{p=0}\\
    &=&-\frac{1}{1-F(0,s)}\frac{\partial F^{*}(p,s)}{\partial p^{}}\bigg|_{p=0}-\frac{F^{*}(0,s)}{\left[1-F(0,s)\right]^{2}}\frac{\partial F^{}(p,s)}{\partial p}\bigg|_{p=0}
\end{eqnarray}
where $F(0,\tau)=\varphi(\tau)$ and $F^*(0,\tau)=\varphi^*(\tau)$. Hence,
\begin{eqnarray}
    \left\langle \tilde{Z}(s|x_{0})\right\rangle _{r}=\frac{\mathcal{L}\left[\varphi^{*}(t)\left\langle Z(t|x_{0})\right\rangle _{0}\right]}{1-\varphi(s)}+\frac{\mathcal{L}\left[\varphi^{}(t)\left\langle Z(t|x_{0})\right\rangle _{0}\right]}{s\left[1-\varphi(s)\right]}.
    \label{Z1}
\end{eqnarray}
Analogously, taking the second derivative of Eq. \eqref{Qrlt} with respect to $p$, we straightforwardly find
\begin{eqnarray}
 \left\langle \tilde{Z}(s|x_{0})^{2}\right\rangle _{r}&=&\frac{2\mathcal{L}\left[\varphi(t)\left\langle Z(t|x_{0})\right\rangle _{0}\right]}{1-\varphi(s)}\left\langle \tilde{Z}(s|x_{0})\right\rangle _{r}\nonumber\\
 &+&\frac{1}{1-\varphi(s)}\left\{ \mathcal{L}\left[\varphi^{*}(t)\left\langle Z(t|x_{0})^{2}\right\rangle _{0}\right]+\frac{1}{s}\mathcal{L}\left[\varphi(t)\left\langle Z(t|x_{0})^{2}\right\rangle _{0}\right]\right\} .
 \label{Z2}
\end{eqnarray}
Following the steps above, we can obtain the corresponding expressions for higher order moments. 

\subsection{Ergodic properties}
Let us consider a stochastic trajectory $\{x(\tau); 0\leq \tau\leq t,x(0)=x_0\}$, i.e.,  observed from $\tau=0$ up to time $\tau = t$. Consider an observable $\mathcal{O}[x(\tau)]$, a function of the trajectory $x(\tau)$. Since $x(\tau)$ is stochastic in nature, the observable $\mathcal{O}[x(\tau)]$ will also fluctuate between the realizations. An observable of the random walk is said to be ergodic if the ensemble average equals the time average $\left\langle \mathcal{O}\right\rangle =\ensuremath{\overline{\mathcal{O}}}$ in the long-time limit. This means that if $\mathcal{O}[x(\tau)]$ is ergodic then its time average $\ensuremath{\overline{\mathcal{O}}}$ is not a random variable. As a consequence, the limiting PDF (the PDF in the long-time limit) of $\ensuremath{\overline{\mathcal{O}}}$ is a delta function centered on its ensemble average, that is,
\begin{eqnarray}
P(\overline{\mathcal{O}},t\to\infty)=\delta\left(\overline{\mathcal{O}}-\left\langle \mathcal{\overline{O}}\right\rangle \right).
     \label{lpdf}
 \end{eqnarray} 
 At this point, let us define the density $P(x,t)$ as the probability of finding the walker at the point $x$ at time $t$, i.e., it is the propagator.
 If the observable is integrable with respect to the density $P(x,t)$, then the ensemble average is
\begin{equation}
    \left\langle \mathcal{O}[x(t)]\right\rangle =\int_{-\infty}^{\infty}\mathcal{O}[x]P(x,t)dx,
\end{equation}
while the time average of $\mathcal{O}[x(t)]$ is defined as
\begin{equation}
    \ensuremath{\overline{\mathcal{O}[x(t)]}=}\frac{1}{t}\int_{0}^{t}\mathcal{O}[x(\tau)]d\tau.
    \label{tav}
\end{equation}
For non-ergodic observables, since $\overline{\mathcal{O}}$ is random, its variance $\textrm{Var}(\overline{\mathcal{O}})$ is non-zero in the long-time limit. Otherwise, for an ergodic observable $\lim_{t\to\infty}\textrm{Var}(\overline{\mathcal{O}})=0$. Keeping this in mind, one can define the ergodicity breaking parameter EB
in the following way
\begin{eqnarray}
\textrm{EB}=\lim_{t\to\infty}\frac{\textrm{Var}(\overline{\mathcal{O}})}{\left\langle \overline{\mathcal{O}}\right\rangle ^{2}}=\lim_{t\to\infty}\frac{\left\langle \overline{\mathcal{O}}^{2}\right\rangle -\left\langle \overline{\mathcal{O}}\right\rangle ^{2}}{\left\langle \overline{\mathcal{O}}\right\rangle ^{2}}.
  \label{EB}
\end{eqnarray}
For ergodic observables, one should have $\textrm{EB}= 0$. 

In the examples below we consider the observable $\mathcal{O}[x(t)]=U[x(t)]$ so that the time average of the observable is from \eqref{tav}
\begin{eqnarray}
    \overline{\mathcal{O}[x(t)]}=\frac{1}{t}\int_{0}^{t}U[x(\tau)]d\tau=\frac{Z(t)}{t}.
    \label{o}
\end{eqnarray}
and so
\begin{eqnarray}
   \left\langle \overline{\mathcal{O}}\right\rangle =\frac{\left\langle Z(t)\right\rangle }{t},\quad\left\langle \overline{\mathcal{O}}^{2}\right\rangle =\frac{\left\langle Z(t)^{2}\right\rangle }{t^{2}}. 
   \label{oo}
\end{eqnarray}
Finally, from \eqref{EB} the ergodicity breaking parameter can be expressed in terms of the two first moments of the functional
\begin{eqnarray}
    \textrm{EB}=\frac{\left\langle Z(t)^{2}\right\rangle }{\left\langle Z(t)\right\rangle ^{2}}-1
    \label{EB2}
\end{eqnarray}
as $t\to \infty$.
Another quantity of interest that characterizes ergodicity is the PDF of the time averaged observable $ \overline{\mathcal{O}[x(t)]}$ around
its mean for long-times, so we define the dimensionless random variable 
\begin{eqnarray}
    \eta =\lim_{t\to \infty} \frac{ \overline{\mathcal{O}}}{\left\langle \overline{\mathcal{O}}\right\rangle}
\end{eqnarray}
and from \eqref{o} and \eqref{oo}, the relative time averaged observable $U[x(t)]$ is defined by 
\begin{eqnarray}
\eta =\lim_{t\to \infty}\frac{Z(t)}{\left\langle Z(t)\right\rangle }.
    \label{eta}
\end{eqnarray}
Once $P(Z,t|0)$ is known, the PDF of $\eta$ follows from
\begin{eqnarray}
   P(\eta)=P(Z=\eta \left\langle Z(t)\right\rangle,t)\left\langle Z(t)\right\rangle.
   \label{peta}
\end{eqnarray}
It is interesting to note that the variance of $\eta$ is nothing but the ergodicity breaking parameter:
$$
\textrm{Var}(\eta)=\left\langle \eta^{2}\right\rangle -\left\langle \eta\right\rangle ^{2}=\left\langle \eta^{2}\right\rangle -1=\frac{\left\langle Z(t)^{2}\right\rangle }{\left\langle Z(t)\right\rangle ^{2}}-1=\textrm{EB}.
$$
Therefore, for ergodic observables (EB = 0) one has  that the limiting PDF of the time averaged observable is $P(\eta)=\delta (\eta -1) $.

\section{General results}
In this section we derive some results for any stochastic functional of a random walk.

\subsection{Limiting PDF}
We consider a resetting time PDF with finite moments. In this case, the Laplace transform of the PDF is, in the long-time limit ($s\to 0$), $\varphi (s)\simeq 1-\langle t \rangle _R s+...$ where $\langle t \rangle _R$ is the mean resetting time. Expressing the term $\exp(-st)$ as a power series we can write
\begin{eqnarray}
   \mathcal{L}\left[\varphi^{}(t)\left\langle Z(t|x_{0})\right\rangle _{0}\right]=\sum_{n=0}^{\infty}\frac{(-s)^{n}}{n!}\int_{0}^{\infty}t^{n}\varphi(t)\left\langle Z(t|x_{0})\right\rangle _{0}dt. 
   \label{expa}
\end{eqnarray}
Most functionals fulfill $\left\langle Z(t|x_{0})\right\rangle _{0}\sim t^{\mu}$ in the long-time limit. The value of $\mu$ depends on the specific dependence of $U[\cdot]$ on $x(\tau)$ and on the underlying random walk. Then,
\begin{eqnarray}
  \int_{0}^{\infty}t^{n}\varphi(t)\left\langle Z(t|x_{0})\right\rangle _{0}dt\sim\int_{0}^{\infty}t^{n+\mu}\varphi(t)dt=\left\langle t^{n+\mu}\right\rangle _{R}<\infty  
  \label{cond}
\end{eqnarray}
where $\left\langle t^{n+\mu}\right\rangle _{R}$ is the moment of order $n+\mu$ of the ressetting time PDF which is finite for any $n$. In consequence we can approximate the expansion \eqref{expa}
as
$$
\mathcal{L}\left[\varphi(t)\left\langle Z(t|x_{0})\right\rangle _{0}\right]=\int_{0}^{\infty}\varphi(t)\left\langle Z(t|x_{0})\right\rangle _{0}dt+O(s).
$$
Analogously, 
$$
\mathcal{L}\left[\varphi^*(t)\left\langle Z(t|x_{0})\right\rangle _{0}\right]=\int_{0}^{\infty}\varphi^*(t)\left\langle Z(t|x_{0})\right\rangle _{0}dt+O(s).
$$
Then, taking the limit $s\to 0$ to Eq. \eqref{Z1} we find
$$
\left\langle \tilde{Z}(s|x_{0})\right\rangle _{r}\simeq\frac{1}{\langle t\rangle_{R}s^{2}}\int_{0}^{\infty}\varphi(\tau)\left\langle Z(\tau|x_{0})\right\rangle _{0}d\tau
$$
which after Laplace inversion yields
\begin{eqnarray}
    \left\langle Z(t|x_{0})\right\rangle _{r}\simeq\frac{t}{\langle t\rangle_{R}}\int_{0}^{\infty}\varphi(\tau)\left\langle Z(\tau|x_{0})\right\rangle _{0}d\tau.
    \label{Z1a}
\end{eqnarray}
Proceeding analogously with Eq. \eqref{Z2} and considering a power dependence of $\left\langle Z(t|x_{0})^2\right\rangle _{0}$ on time, we obtain
\begin{eqnarray}
\left\langle Z(t|x_{0})^{2}\right\rangle _{r}\simeq\frac{t^{2}}{\langle t\rangle_{R}^{2}}\left[\int_{0}^{\infty}\varphi(\tau)\left\langle Z(\tau|x_{0})\right\rangle _{0}d\tau\right]^{2}.
\end{eqnarray}

 In consequence, by virtue of \eqref{EB2}, we obtain EB $=0$ and the observable $\mathcal{O}[x(t)]=U[x(t)]$ is ergodic, which means that in the long-time limit  
 \begin{eqnarray}
     P(Z,t|x_{0})_{r}=\delta\left(Z-\left\langle Z(t|x_{0})\right\rangle _{r}\right)
     \label{ld}
 \end{eqnarray}
 where $\left\langle Z(t|x_{0})\right\rangle _{r}$ is given by Eq. \eqref{Z1a}. Changing the PDF in Eq. \eqref{ld} to the variable $\eta = Z/\left\langle Z(t|x_{0})\right\rangle _r$ one has
 \begin{eqnarray}
     P(\eta|x_0)_r=\delta (\eta -1).
     \label{ld2}
 \end{eqnarray}
It is worth mentioning that we have been able to find the limiting PDF given by Eq. \eqref{ld2} from the two first moments of $Z$ only due to the ergodicity of the observable $U[x(t)]$ when the resetting PDF has finite moments.
However, the same result can be obtained from the characteristic function under resetting times drawn from a PDF with finite moments. We consider the limit where both $s$ and $p$ are small and comparable. This limit corresponds to the bulk region of the PDF of $Z$. Writing $Q(p,t|x_{0})_{0}$ as a moment-generating function
$$
Q(p,t|x_{0})_{0}=\sum_{n=0}^{\infty}\frac{(-p)^{n}}{n!}\left\langle Z(t|x_{0})_{0}^{n}\right\rangle 
$$
one can see that
$$
\mathcal{L}\left[\varphi(t)Q(p,t|x_{0})_{0}\right]=\sum_{n=0}^{\infty}\frac{(-p)^{n}}{n!}\int_{0}^{\infty}e^{-st}\varphi(t)\left\langle Z(t|x_{0})^{n}\right\rangle _0dt
$$
and analogously 
$\mathcal{L}\left[\varphi^*(t)Q(p,t|x_{0})_{0}\right]$. Considering the limit $p\to 0$ to the above expression, the approximations
$$
\mathcal{L}\left[\varphi(t)Q(p,t|x_{0})_{0}\right]   =\varphi(s)-p\int_{0}^{\infty}e^{-st}\varphi(t)\left\langle Z(t|x_{0})\right\rangle _{0}dt+O(p^{2})
$$
and
$$
\mathcal{L}\left[\varphi(t)^{*}Q(p,t|x_{0})_{0}\right]=\varphi^{*}(s)-p\int_{0}^{\infty}e^{-st}\varphi^{*}(t)\left\langle Z(t|x_{0})\right\rangle _{0}dt+O(p^{2})
$$
hold if
\begin{eqnarray}
 \lim_{s\to0}\int_{0}^{\infty}e^{-s\tau}\varphi(\tau)\left\langle Z(\tau|x_{0})^n\right\rangle _{0}d\tau<\infty
 \label{cond2}
\end{eqnarray}
for any $n\in \mathbb{N}$. This condition is necessary to guarantee that all the coefficients of the expansion in powers of $p$ are finite when taking the limit $s\to 0$. Thus, from Eq. \eqref{Qrlt}
\begin{eqnarray}
   \tilde{Q}(p,s|x_{0})_{r}\simeq\frac{\varphi^{*}(s)}{1-\varphi(s)+p\int_{0}^{\infty}e^{-st}\varphi(t)\left\langle Z(t|x_{0})\right\rangle _{0}dt}=\frac{1}{s+p\mathcal{I}(s)}
    \label{Qra}
\end{eqnarray}
as $p\to 0$, where
\begin{eqnarray}
    \mathcal{I}(s)=\frac{s}{1-\varphi(s)}\int_{0}^{\infty}e^{-st}\varphi(t)\left\langle Z(t|x_{0})\right\rangle _{0}dt.
    \label{I}
\end{eqnarray}
If the resetting times PDF has finite moments, then $\varphi (s)\simeq 1-\langle t \rangle _R s+...$  as $s\to 0$, then condition \eqref{cond2} is always satisfied, as can be shown using the same arguments to find \eqref{cond}. Hence, $\lim_{s\to0} \mathcal{I}(s)=\mathcal{I}$ where
$$
\mathcal{I}=\frac{1}{\langle t\rangle_{R}}\int_{0}^{\infty}\varphi(\tau)\left\langle Z(\tau|x_{0})\right\rangle _{0}d\tau.
$$ 
Finally, performing the double Laplace inversion of \eqref{Qra} with respect to $p$ and $s$ we get
$$
P(Z,t|x_{0})_{r}=\delta(Z-\mathcal{I}t)
$$
which is exactly \eqref{ld} provided that $\left\langle Z(t|x_{0})\right\rangle _{r}\simeq\mathcal{I}t$ from \eqref{Z1a}. In terms of the variable  $\eta =Z/\mathcal{I}t$  it reads $P(\eta|x_{0})_{r}=\delta(\eta-1)$ as Eq. \eqref{ld2}. 

\begin{figure}[h!]
    \centering
    \includegraphics[width=0.5\linewidth]{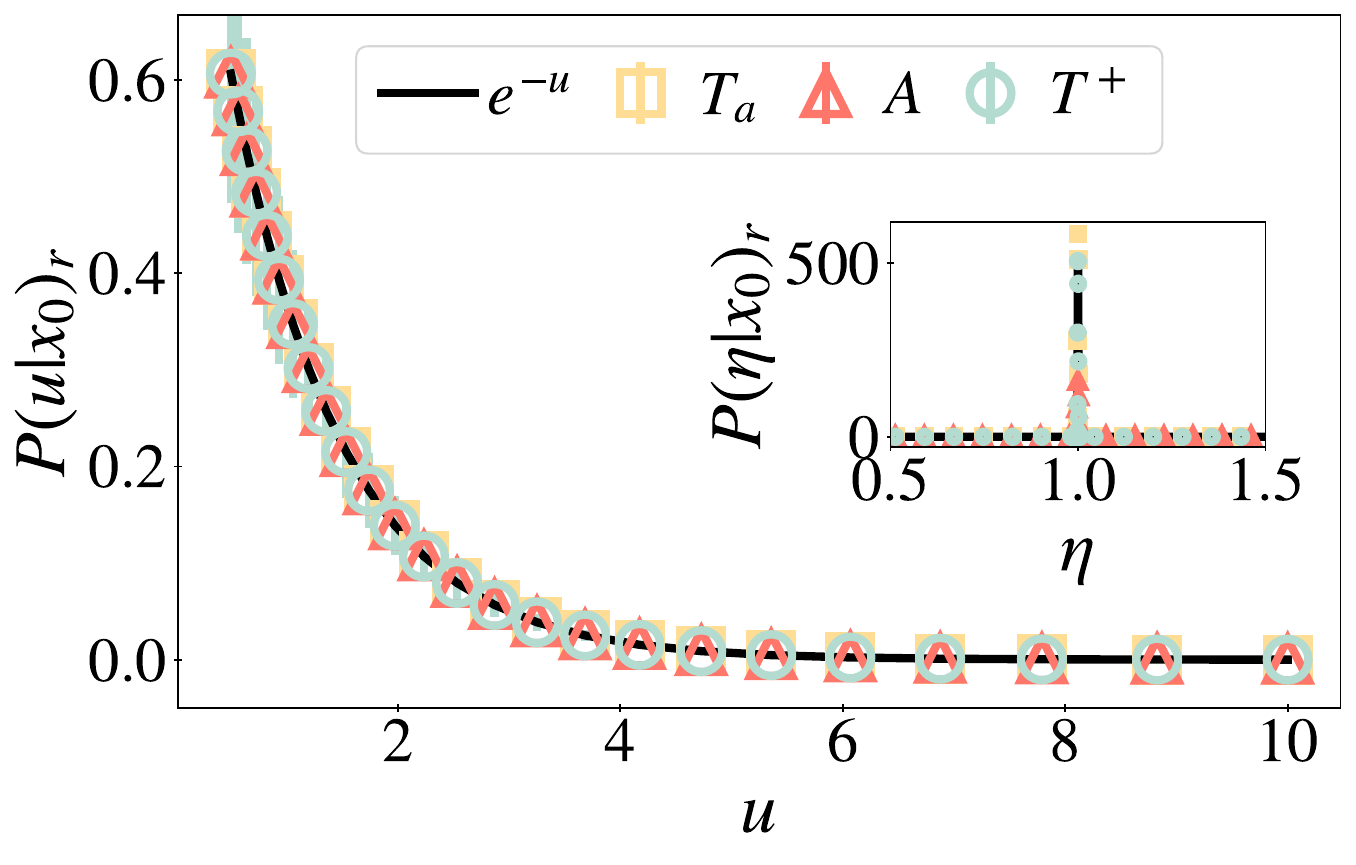}
    \caption{Limiting PDF of observable $Z$ for exponentially distributed resetting of three observables in Laplace space. The black line represents the Laplace transform of Eq. \eqref{ld2} and the symbols are obtained from the Laplace transform of the results of Monte-Carlo simulations. In yellow squares $Z=T_a$ with $a=1$; in orange triangles $Z=A$, and in light-blue circles $Z=T^+$. In the inset we represent the same data in $\eta$ space. For all simulations: $x_0=0$, $D=1$, $\langle t\rangle_R=1.2$ with $\varphi(t)=e^{-t/\langle t\rangle_R}/\langle t\rangle_R$. All simulations are run for $N=10^5$ particles for a total simulation time of $t=10^7$ with a time discretization of $dt=0.1$. The data points are plotted with their error bars.  }
    \label{fig1}
\end{figure}

In Figure \ref{fig1} we show the results for the limiting PDF of three observables: i) the occupation time inside the interval $[-a,a]$, $T_a=\int_0^t \theta (-a<x(\tau)<a)d\tau$, ii) the area under the square position $A=\int_0^t x^2(\tau)d\tau$ and iii) the half occupation time $T^+=\int_0^t \theta (x(\tau))d\tau$, for exponentially distributed resets. We have conducted Monte-Carlo simulations to compute the PDF of the observables when the underlying random walk is a Brownian motion. In our simulations, the time spent by the walker exactly at $x=0$, say $\tau$, is added as $\tau/2$ to the half-occupation time. We compare the data with the theoretical prediction in the Laplace space, in which $P(u|x_0)_r=\int_0^\infty e^{-u\eta}P(\eta|x_0)_rd\eta=e^{-u}$. As can be seen in the figure, the three functionals $T_a$, $A$, and $T^+$ collapse on the theoretical line, which proves that our theoretical result given in Eq. \eqref{ld} holds for any positive functional.  

\subsection{Moments in the long-time limit}
We can obtain the temporal dependence of the first two moments $\left\langle Z(t|x_{0})\right\rangle _{r}$ and $\left\langle Z(t|x_{0})^2\right\rangle _{r}$ in the long-time limit. To do this, we again assume that $\left\langle Z(t|x_{0})\right\rangle _{0}\sim t^{\mu}$ with $\mu>0$, as $t\to \infty$. Let us consider that the resetting times are drawn from a fat-tailed PDF $\varphi (t)\sim t^{-1-\alpha}$ with $0<\alpha\leq 2$. The moments of this PDF are diverging if $0<\alpha<1$, but the first moment is finite if $1<\alpha<2$. In particular, let us consider that the resetting time PDF follows the fat-tailed density
\begin{eqnarray}
    \varphi (t)=\frac{\alpha t_0^\alpha}{t^{1+\alpha}},\quad t>t_0
    \label{pl}
\end{eqnarray}
and 0 otherwise. Note that the limit $\alpha\to 0$ here corresponds to the absence of resetting because in this limit $\varphi\to 0$ and $\varphi^*\to 1$. To compute the long-time limit we need to consider $s\to 0$ in Eqs. \eqref{Z1} and \eqref{Z2}. To do this we need to deal with integrals of the form (see Chapter 13 of \cite{ab64})
\begin{eqnarray}
    \int_{t_{0}}^{\infty}e^{-st}t^{\beta}dt=t_{0}^{\beta+1}e^{-st_{0}}U\left(1,2+\beta,st_{0}\right)
    \label{K}
\end{eqnarray}
where $U(a,b,z)$ is the Kummer $U$ function. In the limit $st_0\ll 1$ this function can be approximated to find (see appendix for details)
$$
\int_{t_{0}}^{\infty}e^{-st}t^{\beta}dt\sim\left\{ \begin{array}{cc}
\textrm{const}, & \beta<-1\\
s^{-1-\beta}, & \beta>-1.
\end{array}\right.
$$
Hence, the terms in \eqref{Z1} and \eqref{Z2}
follow as
$$
\mathcal{L}\left[\varphi(t)\left\langle Z(t|x_{0})^{n}\right\rangle _{0}\right]\sim\int_{t_{0}}^{\infty}e^{-st}t^{n\mu-1-\alpha}dt\sim\left\{ \begin{array}{cc}
\textrm{const}, & \alpha>n\mu\\
s^{\alpha-n\mu}, & \alpha<n\mu.
\end{array}\right.
$$
and
$$
\mathcal{L}\left[\varphi^{*}(t)\left\langle Z(t|x_{0})^{n}\right\rangle _{0}\right]\sim\int_{t_{0}}^{\infty}e^{-st}t^{n\mu-\alpha}dt\sim\left\{ \begin{array}{cc}
\textrm{const}, & \alpha>1+n\mu\\
s^{\alpha-n\mu-1}, & \alpha<1+n\mu.
\end{array}\right.
$$
Keeping only the dominant terms in \eqref{Z1} and \eqref{Z2} we obtain the results of Table 1. Note that the scaling exponent of the moments depend on the values of the exponents $\alpha$ and $\mu$.

\begin{table}[h!]
\centering
\begin{minipage}{0.45\textwidth}
\centering
\begin{tabular}{|c|c|c|}
\hline 
$\alpha$ & $\left\langle Z(t|x_{0})\right\rangle _{r}$ & $\left\langle Z(t|x_{0})^{2}\right\rangle _{r}$\tabularnewline
\hline 
\hline 
$0<\alpha\leq \mu$ & $\sim t^{\mu}$ & $\sim t^{2\mu}$\tabularnewline
\hline 
$\mu<\alpha\leq1$ & $\sim t^{\alpha}$ & $\sim t^{2\alpha}$\tabularnewline
\hline 
$1<\alpha\leq2$ & $\sim t^ {}$ & $\sim t^{2}$\tabularnewline
\hline 
\end{tabular}
\end{minipage}
\hspace{0.05\textwidth}
\begin{minipage}{0.45\textwidth}
\centering
\begin{tabular}{|c|c|c|}
\hline 
$\alpha$ & $\left\langle Z(t|x_{0})\right\rangle _{r}$ & $\left\langle Z(t|x_{0})^{2}\right\rangle _{r}$\tabularnewline
\hline 
\hline 
$0<\alpha\leq 1$ & $\sim t^{\mu}$ & $\sim t^{2\mu}$\tabularnewline
\hline 
$1<\alpha\leq \mu$ & $\sim t^{1+\mu-\alpha}$ & $\sim t^{1+2\mu-\alpha}$\tabularnewline
\hline  
\end{tabular}
\end{minipage}
\caption{Left: long-time limit behavior of the two first moments for $0\leq \mu\leq 1$. Right: long-time limit behavior of the two first moments for $1< \mu\leq 2$.}
\label{table}
\end{table}

For the particular case where $Z$ corresponds to the occupation time in an interval $T_a$, in the absence of resetting, i.e., for free Brownian motion, $\left\langle T_{a}(t)\right\rangle _0\sim t^{1/2}$ and $\left\langle T_{a}^{2}(t)\right\rangle_0 \sim t$ and so $\mu=1/2$. The results in Table \ref{table} predict 
$$
\left\langle T_{a}(t)\right\rangle _{r}\sim\left\{ \begin{array}{cc}
t^{1/2}, & 0<\alpha\leq1/2\\
t^{\alpha}, & 1/2\leq\alpha<1
\end{array}\right.,\quad\left\langle T_{a}^{2}(t)\right\rangle _{r}\sim\left\{ \begin{array}{cc}
t, & 0<\alpha\leq1/2\\
t^{2\alpha}, & 1/2\leq\alpha<1
\end{array}\right.
$$
in agreement with Eqs (118) and C15 in Ref. \cite{BaFlMe23}. 

\begin{figure}[ht!]
    \centering
    \includegraphics[width=\linewidth]{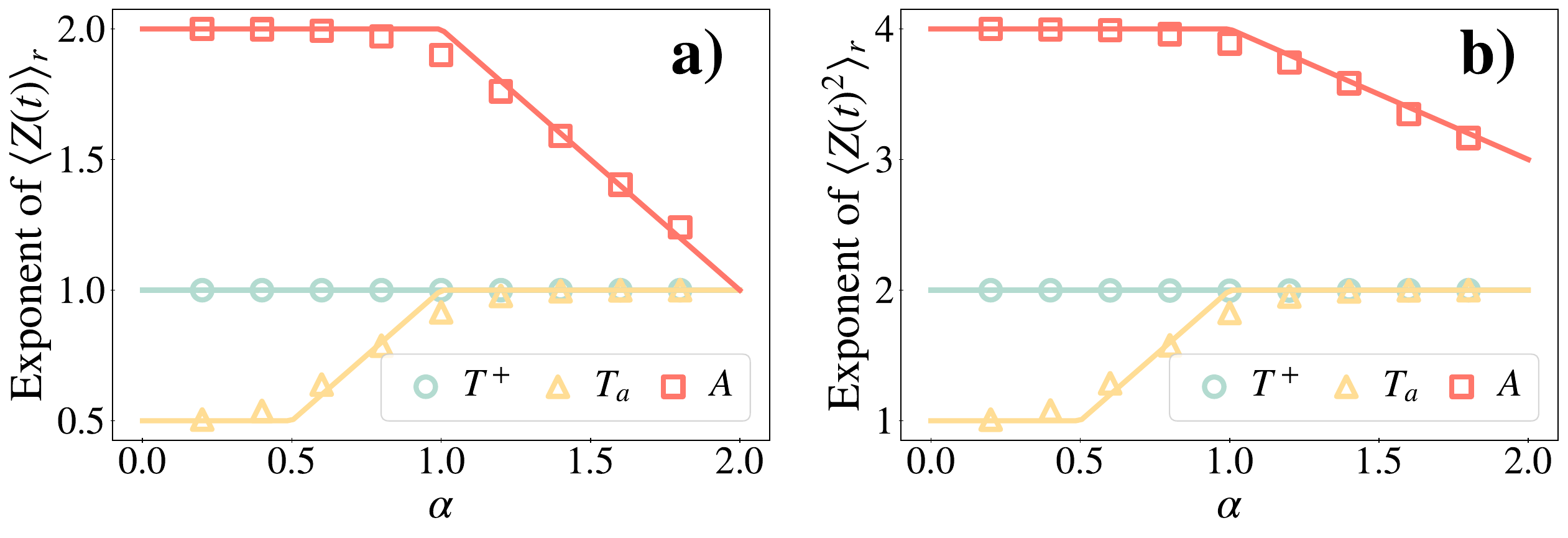}
    \caption{Exponent $\nu$ (of $\sim t^\nu$) of the first (panel a) and second (panel b) moments of the observables $T^+$, $T_a$, and $A$ for a Brownian Walker with fat-tailed resetting (Eq. \eqref{pl}) as a function of the resetting exponent $\alpha$. For every value of $\alpha$, we have computed the values of the observables for the simulation times $t=\{10^3,10^4,10^5,10^6,10^7\}$. Then, we have performed a least squares fit to $\log_{10}(\langle Z(t)^{1,2} \rangle_r)=\nu\log_{10}(t)+c$, the symbols in the plots represent the obtained values of $\nu$. The solid lines are obtained from table \ref{table}. For all simulations, the parameters used are: a time discretization of $dt=0.1$, $D=1$, and $t_0=1$ for the resetting PDF, the points for $\alpha=0.2,1.8$ are run for $N=10^6$ particles, and all the rest for $N=10^5$.}
    \label{fig:exponents}
\end{figure}

In Figure \ref{fig:exponents} we check the results shown in Table \ref{table} for the three different stochastic functionals considered in Figure \ref{fig1}: $T_a$, $T^+$, and $A$. The agreement is excellent.

\section{Application: Half occupation time}
In Ref. \cite{BaFlMe23} we studied the statistical properties of the occupation time in an interval $T_a$, using the backward recurrence time to compute the renewal equation for the PDF of $T_a$. In this section, we apply the previous general results to the case of the half occupation time:
\begin{eqnarray}
    T^{+}(t|x_{0})=\int_{0}^{t}\theta[x(\tau)]d\tau
    \label{T+}
\end{eqnarray}
which is a measure of the time spent by the walker in $x>0$ along the temporal window $[0,t]$. 

\subsection{The two first moments and EB}
In the absence of resetting the mean and the mean square half occupation time of the Brownian motion are $\left\langle T^{+}(t|x_{0})\right\rangle _{0}=t/2$ and $\left\langle T^{+}(t|x_{0})^{2}\right\rangle _{0}=3t^{2}/8$. If we consider the fat-tailed resetting density in Eq. \eqref{pl}, using \eqref{K}, we have 
$$
\mathcal{L}\left[\varphi(t)\left\langle T^{+}(t|x_{0})\right\rangle _{0}\right]=\frac{\alpha t_{0}}{2}e^{-st_{0}}U(1,2-\alpha,st_{0})
$$
and
$$
\mathcal{L}\left[\varphi^{*}(t)\left\langle T^{+}(t|x_{0})\right\rangle _{0}\right]=\frac{1-(1+st_{0})e^{-st_{0}}}{2s^{2}}+\frac{t_{0}^{2}}{2}e^{-st_{0}}U(1,3-\alpha,st_{0})
$$
Analogously,
$$
\mathcal{L}\left[\varphi(t)\left\langle T^{+}(t|x_{0})^{2}\right\rangle _{0}\right]=\frac{3\alpha t_{0}^{2}}{8}e^{-st_{0}}U(1,3-\alpha,st_{0})
$$
and
$$
\mathcal{L}\left[\varphi^{*}(t)\left\langle T^{+}(t|x_{0})^{2}\right\rangle _{0}\right]=\frac{3}{8}\int_{0}^{t_{0}}e^{-st}t^{2}dt+\frac{3t_{0}^{3}}{8}e^{-st_{0}}U(1,4-\alpha,st_{0}).
$$
In the long-time limit we have to consider $st_0\ll 1$ in the previous expressions by using the small argument approximation of the Kummer $U$ function $U(a,b,z)$. Making use of the results derived in appendix A we
thus find
$$
\mathcal{L}\left[\varphi(t)\left\langle T^{+}(t|x_{0})\right\rangle _{0}\right]\sim\frac{\alpha t_{0}}{2}\cdot\left\{ \begin{array}{cc}
\frac{\Gamma(1-\alpha)}{(st_{0})^{1-\alpha}}, & 0<\alpha<1\\
\frac{1}{\alpha-1}, & 1<\alpha<2
\end{array}\right.
$$
$$
\mathcal{L}\left[\varphi^{*}(t)\left\langle T^{+}(t|x_{0})\right\rangle _{0}\right]\sim\frac{t_{0}^{2}}{2}\frac{\Gamma(2-\alpha)}{(st_{0})^{2-\alpha}},\quad0<\alpha<2,
$$
$$
\mathcal{L}\left[\varphi(t)\left\langle T^{+}(t|x_{0})^{2}\right\rangle _{0}\right]\sim\frac{3\alpha t_{0}^{2}}{8}\frac{\Gamma(2-\alpha)}{(st_{0})^{2-\alpha}},\quad0<\alpha<2
$$
and
$$
\mathcal{L}\left[\varphi^{*}(t)\left\langle T^{+}(t|x_{0})^{2}\right\rangle _{0}\right]\sim\frac{3t_{0}^{3}}{8}\frac{\Gamma(3-\alpha)}{(st_{0})^{3-\alpha}},\quad0<\alpha<3.
$$
Finally, from \eqref{Z1}, \eqref{Z2} and the previous results, we obtain
\begin{eqnarray}
    \left\langle T^{+}(t|x_{0})\right\rangle_r \simeq\frac{t}{2},\quad\left\langle T^{+}(t|x_{0})^2\right\rangle_r \simeq\left\{ \begin{array}{cc}
\frac{3-\alpha}{8}t^{2}, & 0<\alpha<1\\
\frac{1}{4}t^{2}, & 1<\alpha<2
\end{array}\right.
\label{eq:moments_tpl}
\end{eqnarray}
in the long-time limit. Now, from \eqref{EB2} the EB parameter easily follows:
\begin{eqnarray}
    \textrm{EB}_{+}=\left\{ \begin{array}{cc}
\frac{1-\alpha}{2}, & 0<\alpha<1\\
0, & 1<\alpha<2
\end{array}\right. .
\label{eq:EB_tpl}
\end{eqnarray}
Note that this result predicts an ergodic transition between non-ergodic and ergodic phases at $\alpha=1$. For $1<\alpha<2$ the observable $\theta [x(\tau)]$ is ergodic, which is in agreement with the fact that for these values of $\alpha$ the resetting times PDF has the first moment finite  and the limiting PDF for $T^+$ converges to the delta function. Note that the condition that the resetting time PDF has to have finite moment can be actually relaxed to require that only the first moment must be finite in this case. For $0<\alpha<1$ the resetting times PDF has infinite moments, the observable is non-ergodic, and the limiting PDF converges towards a \textit{generalized arcsine law} as we show below.  Note that for $\alpha=0$ one has EB$_+=1/2$, which corresponds to the value of EB$_+$ in the absence of resetting. 

\begin{figure}[t!]
    \centering
    \includegraphics[width=0.8\linewidth]{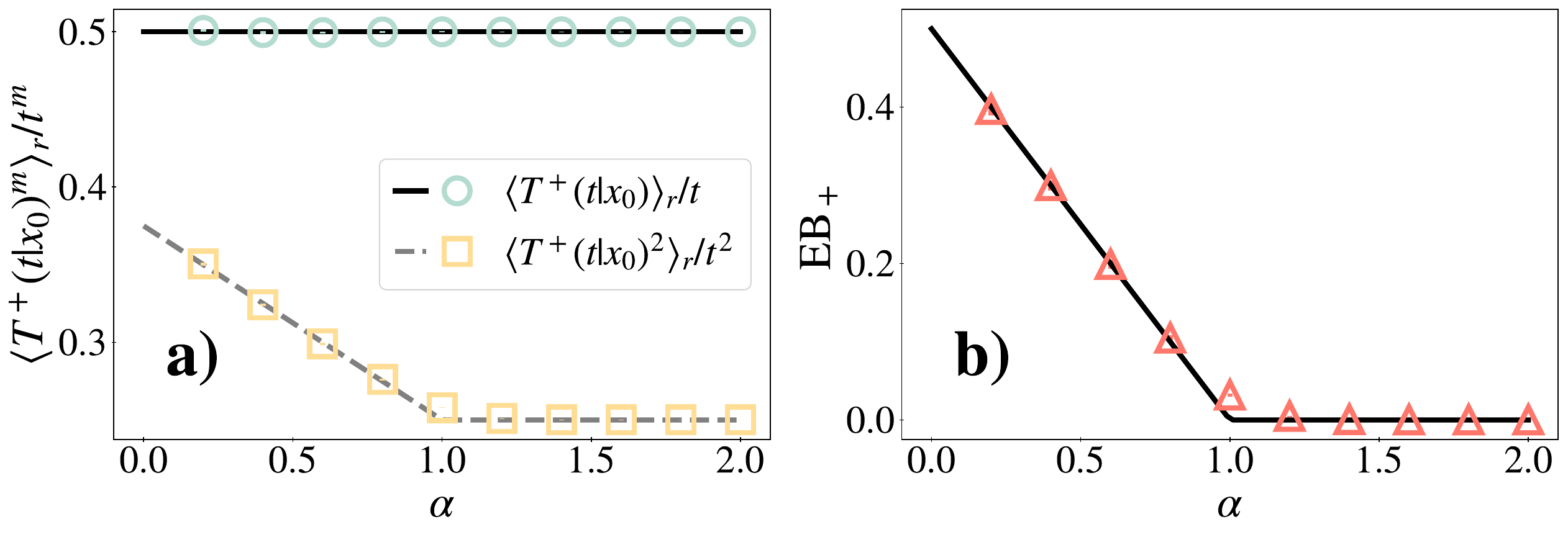}
    \includegraphics[width=0.8\linewidth]{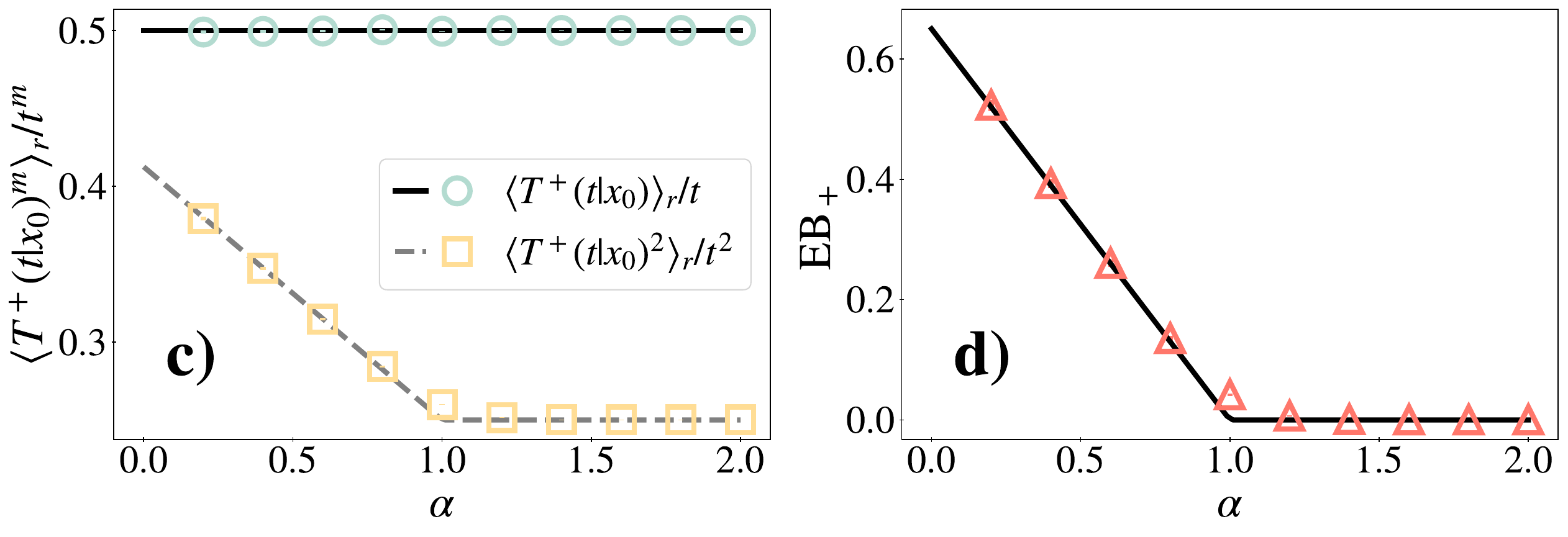}
    \caption{a) Mean ($m=1$) and mean square ($m=2$) half occupation time of a Brownian Motion with fat-tailed resetting (Eq.\eqref{pl}) over the $m$-th power of the total simulation time $t$ as a function of the resetting exponent $\alpha$. The solid black line and the dashed gray line represent Eq.\eqref{eq:moments_tpl} over $t$ and $t^2$ respectively. The symbols are obtained from Monte-Carlo simulations, the light-blue circles represent the mean half occupation time over $t$; and the yellow squares represent the mean square half occupation time over $t^2$. b) The EB parameter for $T^+$ as a function of the resetting exponent $\alpha$. The solid black line represents Eq.\eqref{eq:EB_tpl} and the orange triangles are obtained from the same Monte-Carlo simulations used in panel a). c), and d) Represent the same analysis but for a subdiffusive walker with fat-tailed resetting. In these figures the lines correspond to \eqref{eq:moments2_tpl} in c), and Eq.\eqref{eq:EB2_tpl} in d). The symbols are obtained from Monte-Carlo simulations of a CTRW. For all simulations, the parameters used are: $x_0=0$, and $t_0=1$ for the resetting PDF. For a) and b) $D=1$, and $dt=0.1$ and for c) and d) the waiting time PDF has $t_{\gamma,0}=1$, and $\gamma=0.7$, and the jump PDF $\sigma=0.01$. All simulations are run for $N=10^5$ particles for a total simulation time of $t=10^7$. The data points are plotted with their error bars.}
    \label{fig3}
\end{figure}

It is interesting to note that the presence of resetting reduces the value of EB$_+$ to such an extent that if the resetting is frequently enough (in particular, if $\alpha>1$) the observable becomes ergodic. In Figure \ref{fig3} a) and b) we compare \eqref{eq:moments_tpl} and \eqref{eq:EB_tpl} with numerical simulations. The agreement is excellent.

Let us now consider that the walker moves subdiffusively. In this case, for a Brownian motion without resetting the two first moments of the half occupation time are $\left\langle T^{+}(t|x_{0})\right\rangle _{0}=t/2$ and $\left\langle T^{+}(t|x_{0})^{2}\right\rangle _{0}=(4-\gamma)t^{2}/8$, where $\gamma$ is the exponent of the power-law PDF of waiting times of the walker \cite{MeFlPa25}. Considering the fat-tailed resetting PDF given in Eq. \eqref{pl} then, using \eqref{K}, we have 
\begin{eqnarray}
\left\langle T^{+}(t|x_{0})\right\rangle _{r}\simeq\frac{t}{2},\quad\left\langle T^{+}(t|x_{0})^{2}\right\rangle _{r}\simeq\left\{ \begin{array}{cc}
\frac{4-\gamma-\alpha(2-\gamma)}{8}t^{2}, & 0<\alpha<1\\
\frac{1}{4}t^{2}, & 1<\alpha<2
\end{array}\right.
\label{eq:moments2_tpl}
\end{eqnarray}
in the long-time limit. Now, from \eqref{EB2} the EB parameter easily follows:
\begin{eqnarray}
\textrm{EB}_{+}=\left\{ \begin{array}{cc}
\frac{(1-\alpha)(2-\gamma)}{2}, & 0<\alpha<1\\
0, & 1<\alpha<2
\end{array}\right..
\label{eq:EB2_tpl}
\end{eqnarray}
For $\gamma=1$ the walker moves diffusively and \eqref{eq:EB2_tpl} reduces to \eqref{eq:EB_tpl} as expected. In addition, the transition to the ergodic phase is determined by the values of $\alpha$ for which the resetting PDF has finite first moment, i.e., it occurs again at $\alpha=1$. We proceed in an analogous way than for the diffusive case, and in Figure \ref{fig3} c) and d) we compare the predictions from Eq.\eqref{eq:moments2_tpl} and Eq.\eqref{eq:EB2_tpl} with Monte-Carlo numerical simulations. For the subdiffusive walker, we have simulated a Continuous Time Random Walk (CTRW) with a jump PDF of $\Psi(x) = \frac{\delta(x+\sigma)}{2}+\frac{\delta(x-\sigma)}{2}$ and a power-law waiting time PDF of parameter $\gamma$: $\varphi(t)=\gamma t_{\gamma,0}^\gamma t^{-1-\gamma}, \quad t>t_{\gamma,0}$ and 0 otherwise, $t_{\gamma,0}=1$. Again, we see a good agreement between the analytic predictions and the simulations.

\subsection{Limiting PDF}
To compute the limiting PDF of $Z$ in the presence of resetting, we start from its characteristic function $\tilde{Q}(p,s|x_0)_r$, which can be computed from Eq. \eqref{Qrlt}. Assume that in the long-time limit the PDF $P(Z,t|x_0)_0$ has a scaling solution 
$$
\lim_{t\to\infty}P(Z,t|x_{0})_{0}=\frac{1}{t^{\beta}}f\left(\frac{Z}{t^{\gamma}}\right).
$$
Then, the Laplace transform in the numerator of Eq. \eqref{Qrlt} can be written as
\begin{eqnarray}
  \mathcal{L}\left[\varphi(t)Q(p,t|x_{0})\right]&=&\int_{0}^{\infty}dt\;e^{-st}\varphi(t)\int_{0}^{\infty}e^{-pZ}\frac{1}{t^{\beta}}f\left(\frac{Z}{t^{\gamma}}\right)dZ \nonumber\\
  &=&\int_{0}^{\infty}du\;f(u)\int_{0}^{\infty}e^{-st-put^{\gamma}}t^{\gamma-\beta}\varphi(t)dt  
\end{eqnarray}
where we have introduced the new variable $u=Z/t^{\gamma}$. Note that the integral over time can be expressed as the Laplace transform of $\varphi(t)$ only if $\gamma=\beta=1$. This is the case of the half occupation time, for which the PDF $P(T^+,t|x_0)_0$ has a limiting density of the form $\lim_{t\to\infty}P(T^+,t|x_{0})_{0}=f\left(T^+/t\right)/t$ for $0<T^+<t$. Two relevant examples are the L\'evy asrcsine \cite{Le40} or the Lamperti \cite{La58}  distributions which correspond to case where the underlying random walk is either normal diffusion or subdiffusion, respectively. In this case, $u=T^+/t$ and the Laplace transforms in Eq. \eqref{Qrlt} read
\begin{eqnarray}
   \mathcal{L}\left[\varphi(t)Q(p,t|x_{0})\right]&=&\int_{0}^{1}f(u)\tilde{\varphi}(s+pu)du \nonumber\\
   \mathcal{L}\left[\varphi^*(t)Q(p,t|x_{0})\right]&=&\int_{0}^{1}f(u)\tilde{\varphi}^*(s+pu)du.
   \label{tl}
\end{eqnarray}
To proceed further, we consider the bulk approximation where $s$ and $p$ are both small and comparable. Since $u\in[0,1]$, the argument $s+pu$ is always small in the bulk approximation. Let us consider that the PDF of resettings has a power law fall-off with index $\alpha$
\begin{eqnarray}
    \varphi (t)\sim t^{-1-\alpha},\quad 0<\alpha<2
    \label{pl2}
\end{eqnarray}
as $t\to \infty$. It is not difficult to show that for $0<\alpha<1$ all the moments of $\varphi$ are divergent, while if $1<\alpha<2$, only the first moment $\left\langle t_R\right\rangle =\int_0^{\infty}t\varphi(t)dt$ is finite. Using the Tauberian theorem \cite{Feller}, Eq.~\eqref{pl2} admits the following expansion in Laplace space for small $s$
\begin{eqnarray}
   \tilde{\varphi}(s)=\left\{ \begin{array}{ll}
1-b_{\alpha}s^{\alpha}+..., & 0<\alpha<1\\
1-\left\langle t_R\right\rangle s+..., & 1<\alpha<2
\end{array}\right. 
\label{FI}
\end{eqnarray}
where
$b_{\alpha}$ depends on the details of the specific expression for $\varphi (t)$. With the help of \eqref{FI}, Eqs. \eqref{tl} can be approximated as
\begin{eqnarray}
\mathcal{L}\left[\varphi(t)Q(p,t|x_{0})\right]&\approx&\left\{ \begin{array}{cc}
1-b_{\alpha}\int_{0}^{1}f(u)(s+pu)^{\alpha}du, & 0<\alpha<1\\
1-\left\langle t_R\right\rangle \int_{0}^{1}f(u)(s+pu)du, & 1<\alpha<2
\end{array}\right.\nonumber\\
\mathcal{L}\left[\varphi^{*}(t)Q(p,t|x_{0})\right]&\approx&\left\{ \begin{array}{cc}
b_\alpha\int_{0}^{1}f(u)(s+pu)^{\alpha-1}du, & 0<\alpha<1\\
\left\langle t_R\right\rangle , & 1<\alpha<2
\end{array}\right.
    \label{TL2}
\end{eqnarray}
in the bulk region.

Let us first consider the case $1<\alpha<2$. Plugging \eqref{TL2} into \eqref{Qrlt}
$$
Q(p,s|x_{0})_{r}\approx\frac{1}{\int_{0}^{1}f(u)(s+pu)du}=\frac{1}{s+p\left\langle u\right\rangle }
$$
where $\left\langle u\right\rangle =\int_{0}^{1}uf(u)du$. Since $\left\langle u\right\rangle =\left\langle T^{+}\right\rangle /t$, the double Laplace transform of $Q(p,s|x_{0})_{r}$ with respect to $p$ and $s$ yields
 \begin{eqnarray}
     P(T^+,t|x_{0})_{r}=\delta\left(T^+-\left\langle T^+(t|x_{0})\right\rangle _{r}\right)
     \label{ld3}
 \end{eqnarray}
which is in agreement with \eqref{ld}. 

Next, we consider the case $0<\alpha<1$. By setting \eqref{TL2} in \eqref{Qrlt} we readily obtain
\begin{eqnarray}
   Q(p,s|x_{0})_{r}\approx\frac{\int_{0}^{1}f(u)(s+pu)^{\alpha-1}du}{\int_{0}^{1}f(u)(s+pu)^{\alpha}du}. 
   \label{cf}
\end{eqnarray}
It is worth mentioning that this result establishes a link between the characteristic function of $T^+$ under a resetting with the limiting PDF of $T^+$ in absence of resetting assuming that the resetting PDF decays as $t^{-1-\alpha}$ for $0<\alpha<1$ as $t\to \infty$, regardless of the other details and structure of the resetting PDF. The characteristic function \eqref{cf} can be rewritten in the scaling form 
$$
Q(p,s|x_{0})_{r}\approx \frac{1}{s}g_{\alpha}\left(\frac{p}{s}\right)
$$
where
\begin{eqnarray}
    g_{\alpha}(\chi)=\frac{1}{\chi}\frac{\int_{0}^{1}f(u)(\chi^{-1}+u)^{\alpha-1}du}{\int_{0}^{1}f(u)(\chi^{-1}+u)^{\alpha}du}
    \label{gc}
\end{eqnarray}
and $\chi=p/s$. The double Laplace inversion from $p$ and $s$ to $T^+$ and $t$ has the form $P(T^+,t|x_0)_r\equiv P(z)_r$ where $z=T^+/t$. To find $P(z)_r$  we follow the method in Ref. \cite{GoLu01} by noting
\begin{eqnarray}
P(z)_r=-\frac{1}{\pi z}\lim_{\epsilon\rightarrow0}\textrm{Im}\left[ g_{\alpha}\left(-\frac{1}{z+i\epsilon}\right)\right].
\label{limitd}
\end{eqnarray}
To compute the imaginary part in Eq. \eqref{limitd} we set $\chi^{-1}=-z-i\epsilon$ into the integrals of Eq. \eqref{gc} and rewrite them as
\begin{eqnarray*}
\int_{0}^{1}f(u)(\chi^{-1}+u)^{\alpha-1}du&=&\int_{0}^{1}f(u)(\epsilon^{2}+(u-z)^{2})^{\frac{-1+\alpha}{2}}\cos[(-1+\alpha)\theta(z,u,\epsilon)]du\nonumber\\
&+&i\int_{0}^{1}f(u)(\epsilon^{2}+(u-z)^{2})^{\frac{-1+\alpha}{2}}\sin[(-1+\alpha)\theta(z,u,\epsilon)]du
\end{eqnarray*}
and
\begin{eqnarray*}
 \int_{0}^{1}f(u)(\chi^{-1}+u)^{\alpha}du&=&\int_{0}^{1}f(u)(\epsilon^{2}+(u-z)^{2})^{\frac{\alpha}{2}}\cos[\alpha\theta(z,u,\epsilon)]du\nonumber\\&+&i\int_{0}^{1}f(u)(\epsilon^{2}+(u-z)^{2})^{\frac{\alpha}{2}}\sin[\alpha\theta(z,u,\epsilon)]du
\end{eqnarray*}
where
$$
\theta(z,u,\epsilon)=\tan^{-1}\left(\frac{-\epsilon}{u-z}\right).
$$
Since the integration variable $u$ ranges from 0 to 1 the angle $\theta(z,u,\epsilon)$ may take different values in the limit $\epsilon \to 0$. In particular, if $u\in [0,z]$ then $\lim_{\epsilon\to 0}\theta (z,u,\epsilon)=\pi$ while if $u\in[z,1]$ then $\lim_{\epsilon\to 0}\theta (z,u,\epsilon)=2\pi$. Consequently, the above integrals have to be split in performing the limit $\epsilon \to 0$. Taking the imaginary part of \eqref{gc} using the previous integrals we find
$$
\textrm{Im}g_{\alpha}(z,\epsilon)=\frac{A_{-1+\alpha}\left(zB_{\alpha}-\epsilon A_{\alpha}\right)-B_{-1+\alpha}(zA_{\alpha}+\epsilon B_{\alpha})}{A_{\alpha}^{2}+B_{\alpha}^{2}}
$$
where
$$
A_{\beta}=\int_{0}^{1}f(u)[\epsilon^{2}+(u-z)^{2}]^{\frac{\beta}{2}}\cos[\beta\theta(z,y,\epsilon)]du
$$
and
$$
B_{\beta}=\int_{0}^{1}f(u)[\epsilon^{2}+(u-z)^{2}]^{\frac{\beta}{2}}\sin[\beta\theta(z,u,\epsilon)]du.
$$
Next, we perform the limit $\epsilon \to 0$ to the expressions for $A_{\beta}$ and $B_\beta$. We thus find
$$
\lim_{\epsilon\to0}\textrm{Im}g_{\alpha}(z,\epsilon)=z\frac{\mathcal{A}_{-1+\alpha}\mathcal{B}_{\alpha}-\mathcal{B}_{-1+\alpha}\mathcal{A}_{\alpha}}{\mathcal{A}_{\alpha}^{2}+\mathcal{B}_{\alpha}^{2}}
$$
where
\begin{eqnarray*}
   \mathcal{A}_{\beta}&=&C_{\beta}(z)\cos(\beta\pi)+D_{\beta}(z)\cos(2\beta\pi)\\
\mathcal{B}_{\beta}&=&C_{\beta}(z)\sin(\beta\pi)+D_{\beta}(z)\sin(2\beta\pi)
\end{eqnarray*}
and
\begin{eqnarray}
 C_{\beta}(z)&=&\int_{0}^{z}f(u)(z-u)^{\beta}du=z^{\beta +1}\int_0^1f(zv)(1-v)^\beta dv\nonumber\\
 D_{\beta}(z)&=&\int_{z}^{1}f(u)(u-z)^{\beta}du
 \label{C}
\end{eqnarray}
where we have introduced the new variable $v=u/z$. If the underlying random walk is isotropic then $f(u)=f(1-u)$ and thus $D_\beta (z)=C_\beta (1-z)$.  Finally,  
\begin{eqnarray}
P(z)_{r}=\frac{\sin(\alpha\pi)}{\pi}\frac{C_{\alpha}(z)C_{-1+\alpha}(1-z)+C_{-1+\alpha}(z)C_{\alpha}(1-z)}{C_{\alpha}(z)^{2}+2C_{\alpha}(z)C_{\alpha}(1-z)\cos(\alpha\pi)+C_{\alpha}(1-z)^{2}}.
\label{ldf}
\end{eqnarray}
with $z=T^+/t$. Notably, this is the limiting PDF for $T^+/t$ under resetting times sampled from a PDF with a tail that decays as $t^{-1-\alpha}$ with $0<\alpha<1$. This expression is general and holds for any underlying isotropic random walk. It can be explicitly found if the limiting PDF of $T^+/t$  without resetting is known. It is important to stress that Eq. \eqref{ldf} holds when the resetting time PDF decays as a power law with an exponent such that all moments are divergent, this is, for $\alpha<1$. For $\alpha\geq 1$ the resetting times PDF has the first moment finite  and, therefore, the limiting PDF is not given by \eqref{ldf} but by \eqref{ld}.

It is interesting to analyze the shape of the limiting PDF $P(z)_r$ in terms of the exponent $\alpha$ and other parameters related to the underlying random walk. As we show below, the limiting PDF $P(z)_r$ may have a $\cup$ shape, a W shape, or a $\cap$ shape, depending on the values of $\alpha$. The transition form the $\cup$ shape to the W shape depends on the existence of a maximum at $T^+=t/2$ ($z=1/2$), while the transition from the W shape to the $\cap$ shape depends on the behavior when $T^+$ is close to 0 ($z\to 0^+$) and when $T^+$ is close to $t$ ($z\to 1^-$). To get the equation for the value of $\alpha$ for which  $P(z)_r$ has a local maximum at $z=1/2$ we compute $(d^2P(z)_r/dz^2)_{z=1/2}=0$ using \eqref{ldf} and find
\begin{eqnarray}
C_{\alpha}\left(\frac{1}{2}\right)C_{\alpha-1}^{''}\left(\frac{1}{2}\right)&-&C_{\alpha}^{''}\left(\frac{1}{2}\right)C_{\alpha-1}\left(\frac{1}{2}\right)-2C_{\alpha}^{'}\left(\frac{1}{2}\right)C_{\alpha-1}^{'}\left(\frac{1}{2}\right)\nonumber\\
&=&2\frac{C_{\alpha-1}\left(\frac{1}{2}\right)C_{\alpha}^{'}\left(\frac{1}{2}\right)}{C_{\alpha}\left(\frac{1}{2}\right)}\tan^{2}\left(\frac{\alpha\pi}{2}\right)
\label{a1}
\end{eqnarray}
where the primes stand for derivatives with respect to $z$. This equation has to be numerically solved once the expression for $C_\beta (z)$ is derived from \eqref{C}.

\subsubsection{Brownian walker}

If, for example, the underlying random walk is a Brownian motion then the limiting PDF of $T^+/t$ without resetting obeys the L\'evy arcsine law
\cite{Le40}
\begin{eqnarray}
f(u)=\frac{1}{\pi\sqrt{u(1-u)}}
    \label{as}
\end{eqnarray}
if $0<u<1$ and from \eqref{C} with \eqref{as} we obtain \begin{eqnarray}
C_{\beta}(z)=\frac{\Gamma\left(1+\beta\right)}{\sqrt{\pi}\Gamma\left(\frac{3}{2}+\beta\right)}z^{\frac{1}{2}+\beta}{}_2F_{1}\left(\frac{1}{2},\frac{1}{2};\frac{3}{2}+\beta;z\right),
\label{eq:C_BM}
\end{eqnarray}
where ${}_2F_{1}(\cdot)$ is the Gauss's hypergeometric function. In the limit $u\to 0$,  $f(u)\sim u^{-1/2}$ and from \eqref{eq:C_BM} we see that $C_\beta (z)\sim z^{\beta+1/2}$ as $z\to 0$. Hence, from \eqref{ldf}
\begin{eqnarray}
    P(z)_{r}&\sim& z^{\alpha-1/2},\quad z\to0^{+}\nonumber\\
    P(z)_{r}&\sim& (1-z)^{\alpha-1/2},\quad z\to1^{-}.
    \label{eq:prob}
\end{eqnarray}

As we can see from Eq. \eqref{eq:prob}, when $\alpha>1/2$, the most unlikely values of $T^+$ are 0 and $t$, which are given with vanishing probability. In this case, due to the symmetry of the random walk, the most likely value is $T^+=t/2$, giving rise to a $\cap$ shape limiting PDF.  Then, a sufficiently frequent resetting induces the walker to spend an equal amount of time in the domain $x>0$ and $x<0$, in the long-time limit. Conversely, recall that for the case without resetting, the limiting PDF follows the L\'evy's arcsine law, where the most probable values of $T^+$ are 0 and $t$, implying that a walker is mostly in the positive or negative domain for its entire trajectory. When $\alpha<1/2$, the resetting is less frequent and the probability of $T^+$ at 0 and $t$ does not vanish anymore. In this regime, we can distinguish two possibilities: for a given value of $\alpha$, say $\alpha_c$, if $0<\alpha<\alpha_c$ the limiting PDF has a $\cup$ shape, as in the case without resetting, and the most likely values are $T^+=0$ or $T^+=t$. On the other hand, if $\alpha_c<\alpha<1/2$, the limiting PDF has the W shape and the most likely values are $T^+=0$, $T^+=t$, and $T^+=t/2$, as a transition between $\cup$ and $\cap$ shapes.  The value of $\alpha_c$ can be found by solving Eq. \eqref{a1} with \eqref{eq:C_BM}. The resulting nonlinear equation for $\alpha$ has to be numerically solved. It solution yields $\alpha_c=0.269$.

\begin{figure}[ht!]
    \centering
    \includegraphics[width=0.8\linewidth]{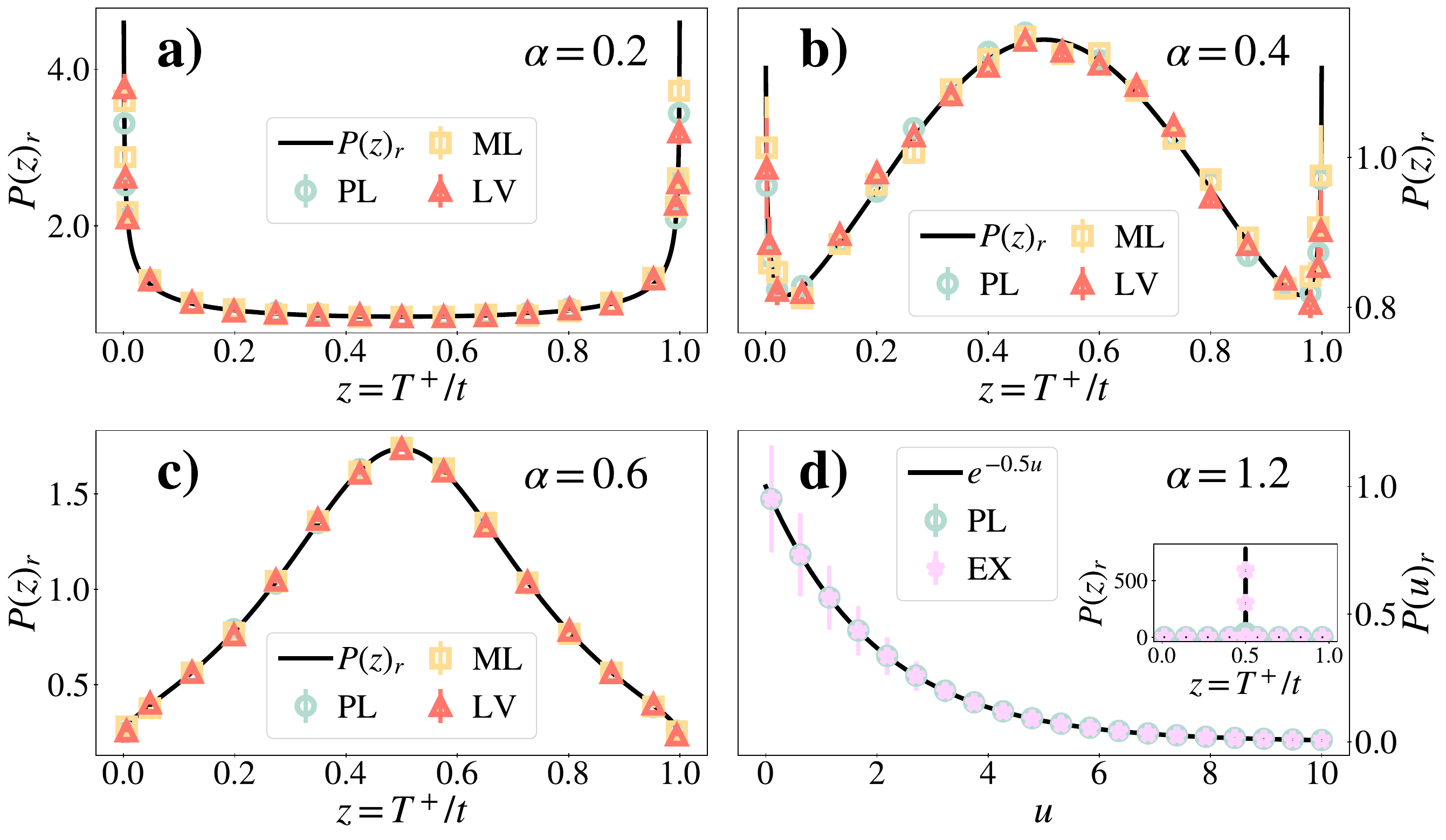}
    \caption{Limiting PDF of the half occupation time of a Brownian Motion with resetting for a variety of resetting distributions. All symbols are obtained from Monte-Carlo simulations of a Brownian walker with different resetting PDFs: power-law (PL) in light-blue circles with $t_0=1$; Mittag-Leffler (ML) in yellow squares; L\'evy  (LV) in orange triangles; and exponential (EX) in pink stars with $\langle t\rangle_R=t_0\alpha(\alpha-1)^{-1}$. In panels a), b), and c) the solid black line represents Eq. \eqref{ldf} with the $C_\beta(z)$ coefficients obtained from Eq. \eqref{eq:C_BM}. In panel d) we present the Laplace transform of Eq. \eqref{ld3} $P(u)_r=\mathcal{L}_{z\to u}[P(z)_r]$. In the inset of panel d) it is plotted in $z$ space. For all simulations the parameters used are: $x_0=0$, $D=1$, and parameter of the resetting distribution, $\alpha=0.2,0.4,0.6$ and $1.2$ for panels a), b), c), and d) respectively. All simulations are run for $N=10^5$ particles for a total simulation time of $t=10^7$ with a time discretization of $dt=0.1$. The data points are plotted with their error bars. }
    \label{fig4}
\end{figure}

In Figure \ref{fig4} we compare Eq. \eqref{ldf} with \eqref{eq:C_BM} against Mont-Carlo simulations conducted for three different resetting time PDFs which decay in time as $t^{-1-\alpha}$ in the long-time limit: power-law (PL) $\varphi(t)=\alpha t_0^\alpha t^{-1-\alpha}, \quad t>t_0$ and 0 otherwise, $t_0=1$; Mittag-Leffler (ML) $\varphi(t)=t^{\alpha-1}\tau^{-\alpha}E_{\alpha,\alpha}(-t^\alpha\tau^{-\alpha})$ with $E_{\alpha,\beta}(\cdot)$ the two-parametric Mittag-Leffler function \cite{Go20,KOZUBOWSKI20011023}, $\tau=\vert \Gamma(1-\alpha) \vert^{1/\alpha}$; and L\'evy (LV), where $\varphi(t) = l_\alpha(t)$ is the one-sided L\'evy stable distribution \cite{Penson2010,WERON1996165}. As can be seen, the limiting PDF \eqref{ldf} agrees with all three cases. This confirms that the limiting PDF depends only on the tail of the resetting times PDF. In addition, we have included the case of exponentially distributed resetting time together with the power-law PDF with $\alpha>1$ (panel d)). In both cases, the resetting times PDF has a finite first moment, and the limiting PDF under resetting fits \eqref{ld2}. It is also interesting to observe the aforementioned transition of the three different shapes W, $\cup$, and $\cap$ for different values of $\alpha$. 

\subsubsection{Subdiffusive walker}

When the underlying random walker performs a subdiffusive motion, then the limiting PDF in the absence of resetting is given by the Lamperti distribution \cite{La58}
\begin{eqnarray}
    f(u)=\frac{\sin(\gamma\pi/2)}{\pi}\frac{\left[u(1-u)\right]^{\frac{\gamma}{2}-1}}{u^{\gamma}+(1-u)^{\gamma}+2\left[u(1-u)\right]^{\frac{\gamma}{2}}\cos(\gamma\pi/2)}.
    \label{lamp}
\end{eqnarray}
Recall that the exponent $\gamma$ is related to the index of fractional derivative, or alternatively, it is the exponent of the waiting time between jumps when its PDF decays as $t^{-1-\gamma}$ with $0<\gamma<1$ in the long-time limit.
In this case, it is not possible to perform analytically the integral in Eq. \eqref{C}. However, we can unveil the behavior of \eqref{ldf} near $z=0$ and $z=1$. This will allow us to know for which value of $\alpha$ and $\gamma$ the transition between the shapes $\cup$ and W appears. Close to $z=0$ the Lamperti distribution \eqref{lamp} behaves as $f(u)\sim u^{\gamma/2-1}$ so that from \eqref{C} we find $C_\beta (z)\sim z^{\beta+1/2}$. Hence,
\begin{eqnarray}
    P(z)_{r}&\sim& z^{\alpha+\frac{\gamma}{2}-1},\quad z\to0^{+}\nonumber\\
    P(z)_{r}&\sim& (1-z)^{\alpha+\frac{\gamma}{2}-1},\quad z\to1^{-}
\end{eqnarray}
Then, if $0<\alpha<\alpha_c$ the PDF $P(z)_r$ has the $\cup$ shape, if $\alpha_c<\alpha<\alpha^*$ it has the shape W where $\alpha^*=1-\gamma/2$, and for $\alpha^*<\alpha<1$ it has the $\cap$ shape. The transition between W and $\cap$ shapes is attained at $\alpha=\alpha_c$ which has to be found as in the previous example. The transition at $\alpha^* = 1-\gamma/2$ can be understood in terms of the First Passage Time (FPT) of the walker to reach the origin. If the FPT has a diverging mean, the walker stays either at $x>0$ or $x<0$, and therefore the limiting PDF of $T^+$ is non-vanishing at 0 and $t$ ($\cup$ or W shapes). On the other hand, when the mean of the FPT is finite, the walker crosses the origin at a finite time, then the probability of the limiting PDF of $T^+$ vanishes at 0 and $t$, i.e., its has a maximum ($\cap$ shape). The transition between a finite and diverging mean for a CTRW with resets is known \cite{MeMa21} and is precisely at $\alpha = 1 - \gamma/2$. The same reasoning is valid for the case of the diffusive walker, as in that case $\gamma = 1$. In Ref. \cite{BoSo20} the authors find a transition at $\alpha=1-\gamma/2$ for the propagator of a subdiffusive walker under a power-law resetting to the initial condition. For sufficiently long times, they find that the propagator changes from being flat near the origin for $\alpha<1-\gamma/2$ to having a sharp peak at $x=0$ with a heavy-tailed decay for larger $x$. This indicates that resets are frequent enough to create strong localization around the reset point $x=0$, while still allowing significant excursions away from it. The resulting profile has no characteristic scale, reflecting the interplay between the underlying random walk and the reset dynamics, in agreement with the transition observed in the limiting PDF of the half-occupation time. This same reasoning can be applied to the case in the previous section of a Brownian walker with power-law resetting, since we have observed a similar transition at $\alpha=1/2$ for the infinite densities of the process \cite{BaFlMe23}.

\begin{figure}[t!]
    \centering
    \includegraphics[width=0.8\linewidth]{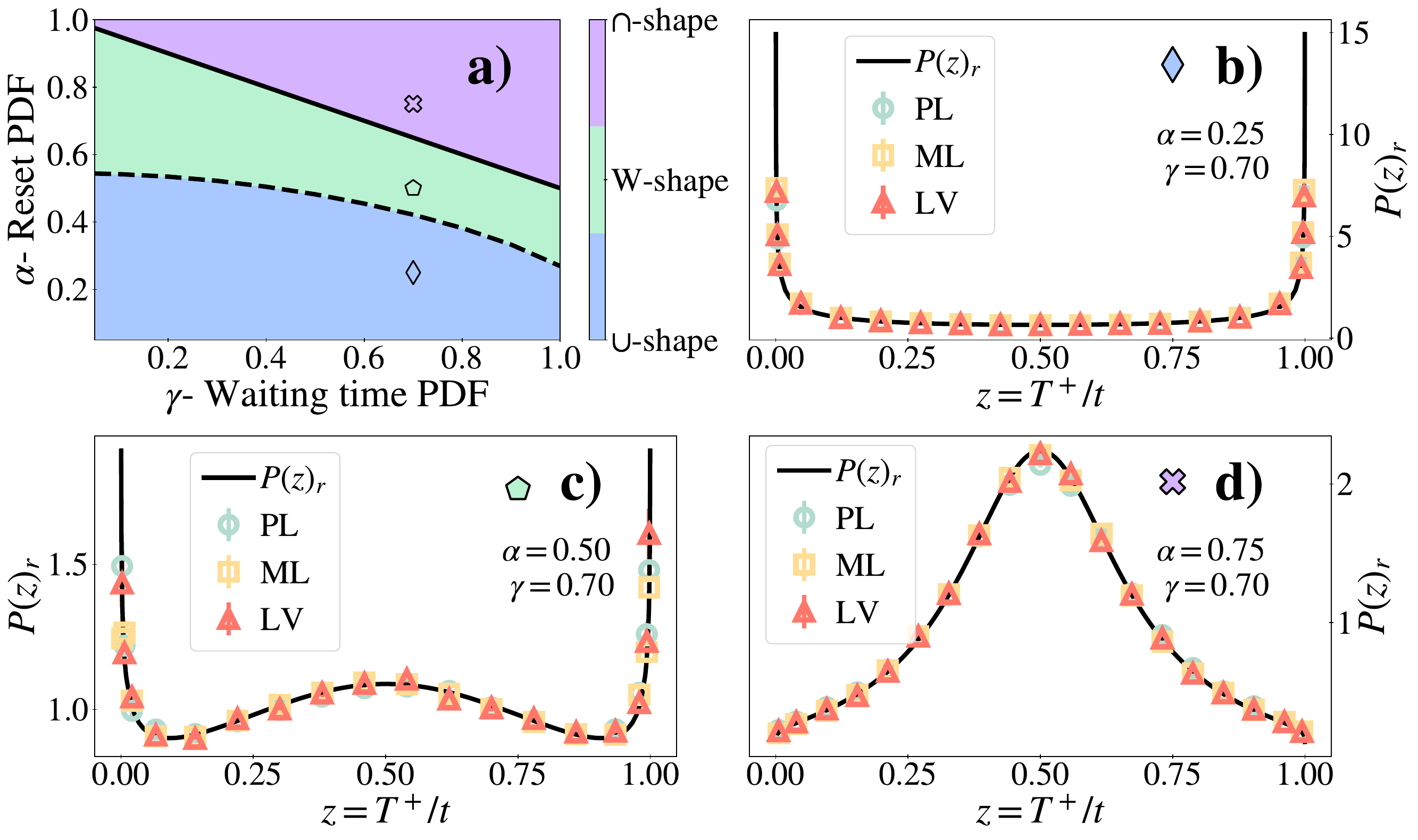}
    \caption{a) Phase diagram of the limiting PDF of $T^+$ of a subdiffusive walker with resetting of exponent $\alpha$. The dashed line represents $\alpha_c$ and the solid line represents $\alpha^*=1-\gamma/2$. The black symbols mark the corresponding $(\gamma,\alpha)$ points represented in panels b), c), and d). b), c), and d) limiting PDF of $T^+$ for a subdiffusive walker with different resetting PDFs: power-law (PL) in light-blue circles with $t_0=1$; Mittag-Leffler (ML) in yellow squares; L\'evy  (LV) in orange triangles. The symbols are obtained from Monte-Carlo simulations of a CTRW with $\sigma=0.01$. The black lines are computed by numerical integration of Eq. \eqref{C} with the $C_\beta(z)$ coefficients obtained from Eq. \eqref{lamp}. For all simulations, the parameters of the resetting distribution are $\alpha=0.25,0.5$, and $0.75$ for panels b), c), and d), respectively. The parameters of the waiting time PDF for all cases are $t_{\gamma,0}=1$, and $\gamma=0.7$. All simulations are run for $N=10^5$ particles for a total simulation time of $t=10^7$, and initial condition $x_0=0$. The data points are plotted with their error bars. }
    \label{fig:subdiff}
\end{figure}

In Figure \ref{fig:subdiff}a) we present the phase diagram for the shapes of $P(z)_r$ in terms of the exponents $\alpha$ and $\gamma$. We also present the dependence of the curves $\alpha^*$ and $\alpha_c$. In the same figure panels b), c), and d) we show a comparison between the curves obtained by numerically integrating Eq. \eqref{C} with the $C_\beta(z)$ coefficients obtained from Eq. \eqref{lamp} and numerical simulations of the CTRW. In the figure, we show an example for the three different cases: $\cup$ shape, W shape, and $\cap$ shape. In all three cases, we obtain a very good agreement for three resetting distributions: power-law, Mittag-Leffler, and L\'evy. This again confirms that the limiting PDF only depends on the tail of the resetting times PDF, also for the subdiffusive case. We want to note that fine numerical precision is needed, especially as $z\to0,1$, to numerically integrate the limiting PDF. This is particularly relevant near the transition values of $\alpha_c$ and $\alpha^*$.


This behavior under resetting is reminiscent of the effects seen when a confining potential is present. Resetting the walker's position to the origin has certain similarities with the presence of a confining potential centered at this point \cite{Evans_2013,Gu24}. Both mechanisms give rise to a non-equilibrium steady state, a finite mean first passage time -depending on the properties of the resetting times PDF and the underlying random walk \cite{MaCaMe19}. They also share some statistical properties of certain stochastic functionals. For example, as we have seen above, the limiting PDF of $T^+$ has the $\cup$ shape depending on the value of $\alpha$. The same happens when the random walker moves in the presence of a force field \cite{Ba06}. The similarity, however, does not extend to the W shape or $\cap$ shape, which do not appear under a confining potential. Nevertheless, the transition between the shapes $\cup$ and W has been already found when a subdiffusive particle moves in a finite symmetric interval \cite{CaBa10} or when a Brownian particle moves in a heterogeneous media \cite{Si22}.

\section{Conclusions}
In this work, we have derived the characteristic function of functionals of random walk when its position is reset to the origin at random times sampled from a general PDF. We have derived the temporal scaling dependence of the first two moments of any stochastic functional of any random walk in the long-time limit when the resetting PDF has a power law tail. 

We have shown that the limiting PDF for any stochastic functional converges to a Dirac Delta function centered at the mean value when the resetting PDF has finite moments.
We have analyzed the ergodic properties of observables and have derived an expression for the EB parameter. In particular, we have computed the EB parameter for the half occupation time of Brownian and subdiffusive walkers under resetting times drawn from a power-law PDF, showing the existence of a transition between non-ergodic and ergodic phases depending on the values of the exponent of the resetting distribution. Additionally, we have seen that for $T^+$, the condition on the resetting PDF can be relaxed to require only a finite first moment for the limiting PDF to converge to a Dirac Delta, and we expect, though have not proved, that this relaxation extends to other functionals whose first moment without resetting scales as  $t^\mu$ with $0<\mu<1$ in the long-time limit.

We have derived a general expression for the limiting PDF of any functional of an isotropic random walk under a power-law resetting PDF if the limiting PDF in absence of resetting  satisfies $\lim_{t\to\infty}P(Z,t|x_0)_0=f(Z/t)/t$. In particular, we have analyzed the half occupation time statistics. We have studied the specific cases of Brownian and subdiffusive underlying random walks and have shown the existence of different shapes of the limiting PDF: $\cup$, W, and $\cap$. Unlike the cases where the walker moves in the presence of a confining potential or in a heterogeneous medium, if it is subjected to a reset whose PDF decays as a power law, a transition between the W and $\cap$ shapes appears. The study of the transition has allowed us to describe the confining role of the reset, which forces the walker to frequently revisit a region close to the origin, and affects the statistical properties of the half occupation time. These insights provide a foundation for broader applications. The general expressions derived in this work can be applied to other functionals of interest, such as the time-averaged position or the area under the trajectory.

\section*{Acknowledgements}
The authors acknowledge the financial support of the Ministerio de Ciencia e Innovaci\'on (Spanish government) under
Grant No. PID2021-122893NB-C22.

\appendix*
\section{Kummer function for small argument}
In this appendix we illustrate the expansion of the Kummer $U$ function (also known as confluent hypergeometric function or simply Tricomi's function) $U(a,b,z)$ is defined as \cite{ab64}
\begin{eqnarray}
    U(a,b,z)=\frac{\Gamma\left(1-b\right)}{\Gamma\left(1+a-b\right)}M(a,b,z)+\frac{\Gamma\left(b-1\right)}{\Gamma\left(a\right)}z^{1-b}M(a+1-b,2-b,z)
    \label{U}
\end{eqnarray}
where
\begin{eqnarray}
    M(a,b,z)=\frac{\Gamma(b)}{\Gamma(a)}\sum_{n=0}^{\infty}\frac{\Gamma\left(a+n\right)}{\Gamma\left(b+n\right)}\frac{z^{n}}{n!}
    \label{M}
\end{eqnarray}
is the Kummer $M$ function. We are interested specifically in the case $a=1$. Inserting the series expansion \eqref{M} into \eqref{U} we can obtain the series expansion for $U(a,b,z)$. Thus, depending on the values of $b$ the leading terms in the limit $z\ll 1$ are 
\begin{eqnarray}
U(1,b,z)=\left\{ \begin{array}{lll}
\frac{1}{1-b}+\frac{z}{b(1-b)}+\dots,\quad & b<0,\\
\frac{1}{1-b}-\frac{\Gamma\left(b\right)}{1-b}z^{1-b}+\dots,\quad & 0<b<1,\\
\frac{\Gamma\left(b-1\right)}{z^{b-1}}+\dots,\quad & b>1.
\end{array}\right.
\label{Us0}
\end{eqnarray}

\bibliography{main}

\end{document}